\documentclass[conference,a4paper]{IEEEtran}
\ifCLASSINFOpdf
\else
\fi
\usepackage{amsmath}
\usepackage{graphicx} 
\usepackage{booktabs}
\usepackage{multirow}
\usepackage{amssymb}
\usepackage{xcolor}
\usepackage{placeins}

\definecolor{ycCoral}{RGB}{255,127,80}

\begin{document}
\title{Decoder-Guided Lossy Contour Coding \\ Via Anchor Refinement\vspace{-1em}}

%\author{Anonymous Authors}
 \author{
 \IEEEauthorblockN{Ruoyu Yang, Yichi Zhang, and Fengqing Zhu}
 \IEEEauthorblockA{Elmore Family School of Electrical and Computer Engineering, Purdue University\\
 \{yang2729, zhan5096, zhu0\}@purdue.edu}
 }

%Telephone: (800) 555--1212, Fax: (888) 555--1212}


\maketitle

\begin{abstract}
%\boldmath
Object contours serve as compact structural priors for many receiver-side vision tasks such as image super-resolution, edge-conditioned generation, and machine vision. When such tasks are deployed over a bandwidth-limited channel, the sender transmits the high-quality object contour as structural side information to guide reconstruction at the receiver, while—to save bandwidth—only a low-quality reference such as a downsampled image or base-layer reconstruction is delivered. As a result, the decoder can already extract a coarse contour from this reference at no transmission cost, creating an encoder–decoder asymmetry: the fine contour must be coded and sent, yet a free coarse version is available at the decoder. This asymmetry is ignored by existing contour codecs such as JBIG2 and chain coding, which are lossless, symmetric, and offer no rate–distortion control, leading to high bitrates. In this paper, we propose a coarse-to-fine contour coding framework that models a high-quality contour as a structured geometric refinement of the decoder-available coarse contour. The encoder extracts ordered anchors along the fine contour and performs adaptive anchor skipping under a distortion constraint. The decoder then reconstructs the contour by using the coarse prior to guide anchor connectivity. This formulation enables lossy contour compression with an explicit rate–distortion trade-off. Experiments show \textbf{\boldmath $54.5\%$--$66.9\%$} bitrate reduction over methods without decoder-side guidance, and up to 5× savings over JBIG2, while preserving high geometric accuracy. 

\end{abstract}



% \fi
\IEEEpeerreviewmaketitle

%\section*{Acknowledgment}
%The authors would like to thank Jianliang Yi, Weiqiang Lei and other colleagues from TCL for their valuable support throughout this work, particularly for their assistance with code implementation, technical advices and computational supports.



%\bibliographystyle{IEEEtran}
%\begin{thebibliography}{1}

  %0.5em minus 0.4em\relax Harlow, England: Addison-Wesley, 1999.

%\end{thebibliography}
%% Build the formatted document
%%
\maketitle

\section{Introduction}
\label{intro}
High-fidelity visual reconstruction often depends not only on texture statistics but also on accurate geometric structure. Structural and edge priors have been exploited in super-resolution and reconstruction tasks~\cite{basicvsr,basicvsrpp,stavsr,egvsr,eocvsr}, and edge-aware video coding~\cite{epdmm}. More recently, metadata-guided workflows have made this structural guidance explicit: the Rich Detail Range (RDR)~\cite{rdr} argues that perceived quality hinges on the richness and fidelity of structural detail rather than luminance and color alone, while MetaSR~\cite{metasr} transmits compact metadata to condition generative super-resolution. Object contours are a particularly compact form of such structural metadata: they preserve object boundaries, shape, and spatial layout while discarding most appearance information, making them attractive side information for super-resolution, frame interpolation, and edge-conditioned generation.

Contour transmission, however, creates a coding problem under tight bitrate
budgets. Classical contour coders form two families. Chain-based methods encode boundaries as pixel-wise directional chains, from Freeman coding~\cite{freeman1961} to differential and entropy-coded variants~\cite{liu2005}, whereas bi-level and context-based arithmetic coders such as JBIG2~\cite{jbig2}, adaptive contour-map CAE~\cite{pinho2001}, and MPEG-4 shape coding~\cite{brady1997} treat the contour as a binary image. Both families are lossless and provide no contour-level rate--distortion (RD) control. A second line of work introduces geometric approximation: polygonal and vertex-based representations~\cite{douglas1973} and, in particular, RD-optimal boundary encoding~\cite{schuster1997,schuster1998,mpeg4shape} trade boundary accuracy for rate. These methods do offer RD control, but they still encode each contour as a standalone signal transmitted in full, independent of any information already available at the receiver.

%\begin{figure}[t]
%\centering
%\includegraphics[width=\linewidth]{sample_anchor.png}
%\caption{
%x4 ISR anchor refinements example.}
%\label{fig:anchor}
%\end{figure}

A key property in many practical workflows is that the encoder and decoder are asymmetric: the decoder already has access to a low-quality reference, such as a low-resolution input or a base-layer reconstruction, from which a coarse contour can be extracted without additional transmission cost. Although this coarse contour lacks fine boundary accuracy, it supplies approximate topology, spatial ordering, and object-level connectivity. Leveraging information already present at the decoder to lower the transmission rate is the guiding principle of source coding with decoder side information~\cite{slepianwolf,wynerziv}; yet existing contour codecs ignore this asymmetry and re-transmit geometry that the decoder could largely infer from its own coarse contour.

In this paper, we propose a decoder-guided coarse-to-fine contour coding
framework. Instead of coding the high-quality contour independently, we model it as a structured geometric refinement of a decoder-available coarse contour. The encoder simulates the decoder-side coarse contour, extracts ordered anchor points from the high-quality contour, and adaptively skips anchors under a point-to-segment distortion constraint. The retained anchors are entropy coded as sparse local refinements along the coarse-contour chain. At the decoder, the coarse contour provides connectivity and ordering, allowing the decoded anchors to reconstruct a fine contour without explicitly transmitting the full contour chain. By adjusting the anchor-skipping threshold, the codec produces different rate--distortion operating points between bitrate and geometric fidelity. To isolate the benefit of decoder-side guidance, we further introduce an HQ-only anchor coding baseline that uses the same selected anchors but transmits their coordinates explicitly, directly quantifying the rate saved by the decoder-available coarse geometry.

The main contributions of this paper are as follows:
\begin{itemize}
\item We formulate decoder-guided lossy contour compression, casting fine-contour coding as refinement of a coarse contour that is freely available at the decoder as side information, with explicit RD control.
\item We propose an anchor-based coarse-to-fine codec with controllable rate--distortion trade-offs, and show its bitrate advantage over HQ-only
anchor coding and JBIG2.
\end{itemize}

%The remainder of this paper is organized as follows. Section~II describes the proposed method in detail, including coarse contour construction, ordered anchor extraction and adaptive skipping, entropy coding design, and the HQ-only baseline. Section~III presents experimental results on a video contour dataset with comparisons against JBIG2 and the HQ-only baseline. Section~IV concludes the paper.
\section{Our Method}
\label{sec:method}

\subsection{Overview of Decoder-Guided Contour Coding}
\label{subsec:overview}

\begin{figure}[t]
\centering
\includegraphics[width=\linewidth]{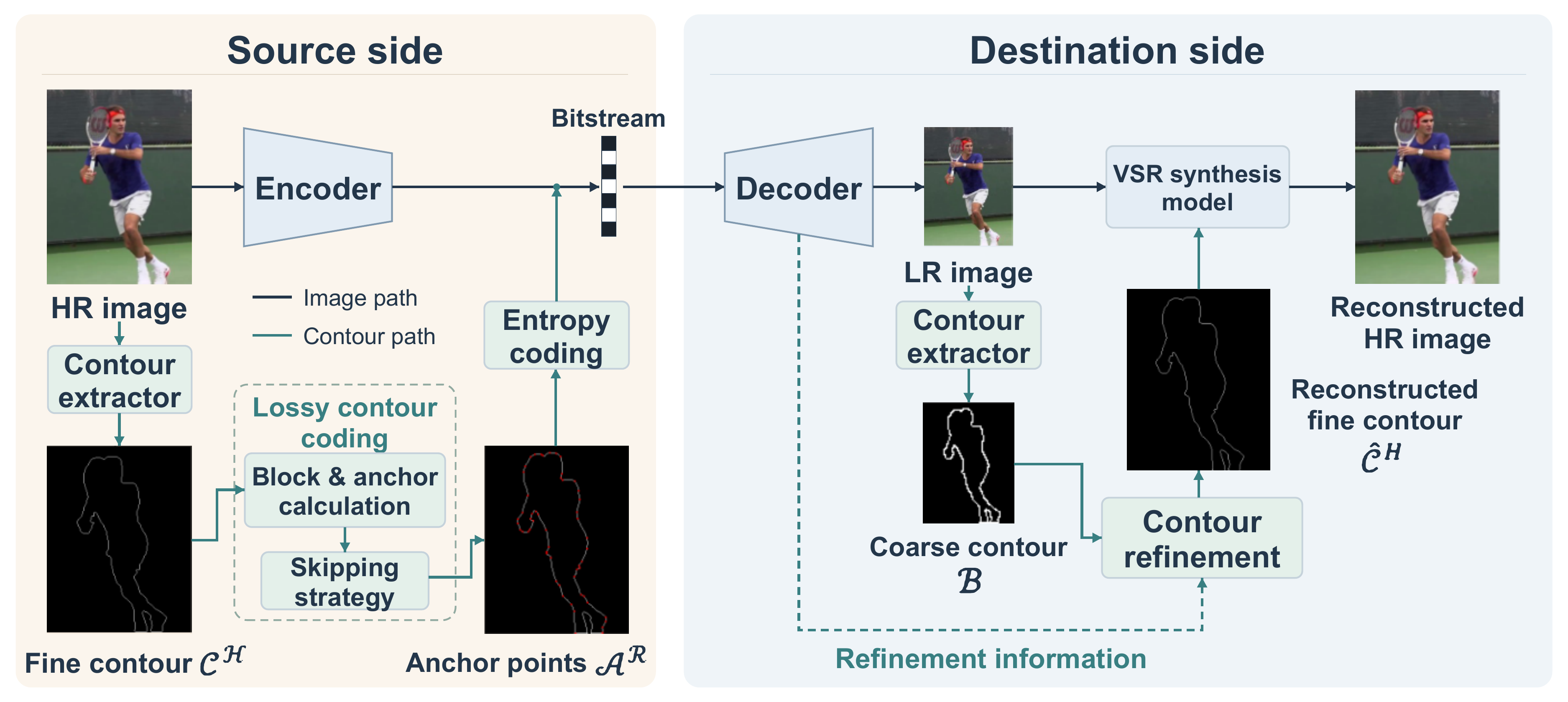}
\caption{
Overview of contour-assisted $\times 4$ VSR with decoder-guided lossy contour coding. The source transmits sparse anchor refinements of the HR contour, while the destination extracts a coarse contour from the LR image and uses the decoded refinements to reconstruct a fine contour for VSR synthesis. This paper focuses on the lossy contour coding and contour reconstruction module}
\label{fig:framework}
\end{figure}

We consider the compression of a high-quality binary contour map $\mathcal{C}^{H}\subset\mathbb{Z}^2$, where foreground pixels represent object boundaries. In decoder-guided vision pipelines, a low-quality reference, such
as a downsampled input or base-layer reconstruction, is already available at
the decoder, from which a coarse contour $\mathcal{C}^{L}$ can be extracted
without extra transmission. As illustrated in Fig.~\ref{fig:framework}, we therefore model $\mathcal{C}^{H}$ as a geometric refinement of $\mathcal{C}^{L}$ rather than coding it as a standalone binary map. The encoder simulates the decoder-side coarse contour, selects sparse anchors on $\mathcal{C}^{H}$, and transmits only their local refinements along the coarse chain. A user-controlled anchor-skipping threshold produces different rate--distortion operating points. At the decoder, the reconstructed contour $\hat{\mathcal{C}}^{H}$ is obtained by placing the decoded anchors on the corresponding coarse contour positions and connecting them according to the coarse-chain order. 

%This formulation differs from conventional contour codecs such as JBIG2 or chain coding. Those methods treat the contour as a standalone binary or pixel-wise chain signal. In contrast, our method exploits decoder-side geometric side information, so that the encoder does not need to explicitly transmit the full contour topology or ordering. The resulting codec supports lossy contour compression with an explicit trade-off between bitrate and geometric fidelity.

\subsection{Ordered Anchor Refinement on Coarse Contours}
\label{subsec:anchor}

Given the high-quality contour $\mathcal{C}^{H}$, we first extract an ordered fine contour
\begin{equation}
    \mathcal{P}^{H} = \{p_1, p_2, \ldots, p_N\},
\end{equation}
where $p_i=(r_i,c_i)$ denotes a contour pixel, with $r_i$ and $c_i$ representing its row and column coordinates in the image, respectively. To construct the side information that is assumed to be available at the decoder, the encoder simulates the decoder-side contour extraction process by downsampling the high-quality contour into a coarse occupancy map. Specifically, the image is divided into non-overlapping $s \times s$ blocks, where $s=4$ in our implementation, and a coarse block is marked as active if it contains at least one fine contour pixel. This step aligns the encoder-side anchor selection with the decoder-side coarse contour prior, so that the transmitted refinement
can be interpreted relative to the same coarse geometric support.

To avoid ambiguous internal paths caused by downsampling and discretization, we keep only the outer boundary of the active-block region and use boundary following to trace it into an ordered block chain, which serves as the block-level representation of the decoder-side coarse contour $\mathcal{C}^{L}$:
\begin{equation}
    \mathcal{B} = \{B_1, B_2, \ldots, B_T\},
\end{equation}
where $B_t$ denotes the $t$-th coarse block and $T$ is the number of coarse-chain positions. For each $B_t$, we collect the fine contour pixels falling inside this block:
\begin{equation}
    \mathcal{S}(B_t) = \{p \in \mathcal{C}^{H}: p \in B_t\}.
\end{equation}
The anchor point associated with $B_t$ is selected as the fine contour pixel closest to the centroid of $\mathcal{S}(B_t)$:
\begin{equation}
    a_t =
    \arg\min_{p \in \mathcal{S}(B_t)}
    \left\|p - \frac{1}{|\mathcal{S}(B_t)|}
    \sum_{q\in\mathcal{S}(B_t)} q \right\|_2 .
\end{equation}
This gives an ordered anchor sequence
$\mathcal{A}=\{a_1,a_2,\ldots,a_T\}$ whose ordering is inherited from the decoder-available coarse contour chain. Equivalently, each anchor $a_t$ is addressed by the index $t$ of its block in the coarse chain $\mathcal{B}$ together with its local position inside that block, i.e., $a_t = o(B_t) + (r_t^{\text{loc}}, c_t^{\text{loc}})$, where $o(B_t)$ is the top-left origin of $B_t$ and $0 \le r_t^{\text{loc}}, c_t^{\text{loc}} < s$. This decomposition underlies the entropy coding of Sec.~\ref{subsec:entropy_multi}: since the decoder
already has access to $\mathcal{B}$, transmitting the block-chain index and
the local position avoids coding the absolute image coordinate, thereby
reducing the bitrate.

To enable rate--distortion control, we apply anchor skipping. For a candidate anchor $a_i$, let $a_i^{-}$ and $a_i^{+}$ be its current retained predecessor and successor in the ordered sequence. If removing $a_i$ introduces only a small local approximation error,
\begin{equation}
    d(a_i, \overline{a_i^{-}a_i^{+}}) \leq \tau ,
\end{equation}
where $d(\cdot,\cdot)$ is the Euclidean distance from a point to a line segment and $\tau$ is a user-controlled skipping threshold, then $a_i$ is removed. The deletion is performed dynamically: once an anchor is removed, the predecessor/successor links of its two neighbors are relinked before testing subsequent anchors. After skipping, the retained anchor sequence is
$\mathcal{A}^{R}=\{a_{i_1},a_{i_2},\ldots,a_{i_M}\}$, where $M$ is the number of retained anchors. The decoder reconstructs the contour by connecting adjacent retained anchors following the coarse-chain
order using Bresenham line rasterization, and closes the loop for closed contours.

\subsection{Entropy Coding and Multi-object Coding}
\label{subsec:entropy_multi}

\begin{figure}[t]
    \centering
    \includegraphics[width=\linewidth]{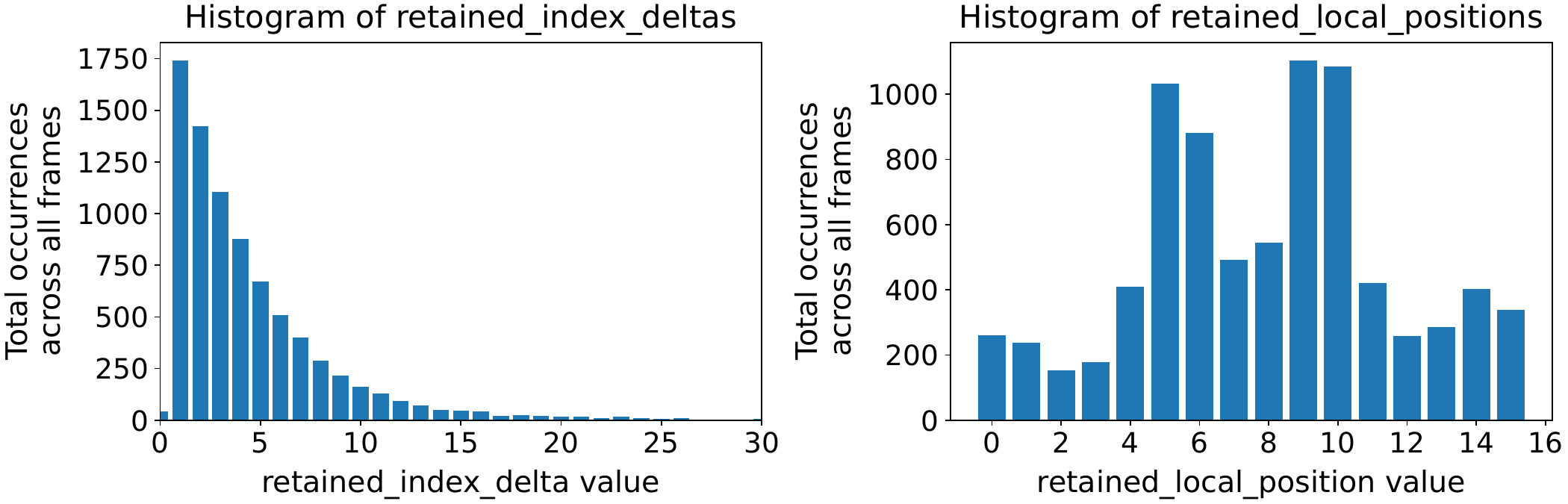}
    \caption{
    Empirical symbol distributions used for entropy coding.
    Left: retained-anchor relative index distribution, which is concentrated
    at small positive values and is suitable for Rice coding.
    Right: anchor local-position distribution inside a $4\times4$ block,
    which is coded using a fixed global Huffman codebook.
    }
    \label{fig:entropy_hist}
\end{figure}

For each object contour, the retained anchors are represented by three syntax elements. First, the number of retained anchors $M$ is transmitted. Second, we encode the retained-anchor indices along the original coarse chain. Let $i_k$ be the original coarse-chain index of the $k$-th retained anchor, with $1\leq i_k \leq T$. Instead of sending absolute indices, we transmit their differences:
\begin{equation}
    \Delta i_1 = i_1,\qquad
    \Delta i_k = i_k - i_{k-1}, \quad k=2,\ldots,M .
\end{equation}
Because retained anchors usually remain close along the ordered coarse chain, the relative indices are positive integers concentrated around small values. As shown in Fig.~\ref{fig:entropy_hist}, their empirical distribution decays rapidly and most values fall within a small range. We therefore use Rice coding for the relative-index stream.

Third, for each retained anchor, we encode its local position inside the corresponding $s\times s$ coarse block. The local position is represented as
\begin{equation}
    l_k = r_k^{\text{loc}} \cdot s + c_k^{\text{loc}},
    \qquad l_k \in \{0,\ldots,s^2-1\}.
\end{equation}
Compared with the relative-index stream, the anchor local positions are more evenly distributed over the block. We  use a global Huffman codebook for local-position coding. The codebook is estimated offline once from the held-out validation set statistics and fixed for both encoder and decoder, so it does not need to be transmitted for each sequence. 
%The final bitrate is computed as the sum of the retained-count bits, Rice-coded relative-index bits, and Huffman-coded local position bits. The coded rate includes all syntax required to decode the transmitted refinement stream.

For multi-object frames, connected contour components are encoded
independently. To synchronize encoder and decoder, objects are sorted
lexicographically by the top-left coordinate of their coarse bounding boxes.
The final reconstruction is obtained by the union of all object-wise decoded
contours.

\subsection{HQ-Only Anchor Coding Baseline}
\label{subsec:hqonly}

To quantify the benefit of decoder-side coarse guidance, we introduce an HQ-only anchor coding baseline. This baseline uses the same encoder-side
anchor extraction and anchor skipping procedure as the proposed method, so the geometric approximation quality is comparable. However, it does not assume
that the decoder has access to an ordered coarse contour chain. Therefore, the bitstream must explicitly describe both the positions and ordering of the retained anchors.

For each object, the HQ-only codec transmits the first retained anchor using its absolute image coordinate and represents all subsequent anchors by a relative coordinate chain:
\begin{equation}
    \Delta a_k = a_{i_k} - a_{i_{k-1}},
    \qquad k=2,\ldots,M .
\end{equation}
Each $\Delta a_k=(\Delta r_k,\Delta c_k)$ is an image-space coordinate delta. The decoder reconstructs the contour solely from this transmitted ordered anchor sequence by connecting consecutive anchors using Bresenham line rasterization. We denote the resulting rate as $R_{\text{HQ-only}}$. The absolute first anchor is coded with fixed-length image coordinates, and the subsequent coordinate deltas are entropy coded using rice coding. Comparing the proposed decoder-guided bitrate against $R_{\text{HQ-only}}$ directly measures the coding gain provided by the decoder-available coarse contour, which supplies contour topology, ordering, and approximate geometry without explicit transmission.
\section{Experiments}
\label{sec:experiments}

\subsection{Dataset and Evaluation Protocol}
\label{subsec:exp_setup}

We construct the evaluation set from SportsSloMo~\cite{slomo2024}, a public human-centric sports video benchmark. Specifically, we use 49 sports clips covering basketball, American football, tennis, and combat sports, with each clip standardized to 32 frames at 8 fps and $1280\times720$ resolution. For each frame, we apply the proposed anchor-based contour coding pipeline to the input contour map. The block size is fixed to $4\times4$ by default (we ablate $s$ in Sec.~\ref{subsec:ablation}).

The high-quality contour $\mathcal{C}^{H}$ is extracted using SAM3, followed by binarization, small-component removal, and 1-pixel thinning. Since this paper focuses on contour coding rather than contour extraction,we emulate the decoder-side $\times4$
coarse contour by projecting $\mathcal{C}^{H}$ onto a $4\times4$ block grid to form the decoder-available coarse contour $\mathcal{C}^{L}$. We then keep the 1-pixel outer boundary of $\mathcal{C}^{L}$ to obtain an ordered coarse chain. We assume the encoder-side simulated coarse contour and the decoder-side coarse contour are synchronized, and only the refinement information is included in the bitrate.

To obtain different rate--distortion operating points, we vary the anchor
skipping threshold as
\begin{equation}
\tau \in \{1.0, 1.2, 1.4, 1.6, 1.8, 2.0\}.
\end{equation}
A larger $\tau$ removes more anchors and therefore reduces the bitrate at the
cost of lower contour accuracy. For each threshold, all frames from all clips are processed and aggregated into one dataset-level RD point.

We measure rate using bits per contour pixel (bpcp), defined over the entire
dataset as
\begin{equation}
\mathrm{bpcp}
=
\frac{\sum_{v}\sum_{t} R_{v,t}}
{\sum_{v}\sum_{t} |\mathcal{C}^{H}_{v,t}|},
\end{equation}
where $R_{v,t}$ is the coded bitstream size of frame $t$ in video $v$, and
$|\mathcal{C}^{H}_{v,t}|$ is the number of input contour pixels. This global
ratio follows the standard video-rate aggregation practice and avoids over-weighting frames with very sparse contours.

For distortion, we report region IoU between the input contour region and the
reconstructed contour region:
\begin{equation}
\mathrm{IoU}_{\mathrm{region}}
=
\frac{
|\Omega(\mathcal{C}^{H}) \cap \Omega(\hat{\mathcal{C}}^{H})|
}{
|\Omega(\mathcal{C}^{H}) \cup \Omega(\hat{\mathcal{C}}^{H})|
},
\end{equation}
where $\Omega(\cdot)$ denotes the filled region induced by a contour. Higher IoU indicates better geometric fidelity.

\subsection{Compared Methods}
\label{subsec:compared_methods}

We compare three coding schemes.

\textbf{Proposed decoder-guided codec.}
This is our coarse-to-fine contour codec. The decoder-side coarse contour
provides the coarse chain. The bitstream
only encodes the retained-anchor count, retained-anchor relative indices along
the coarse chain, and anchor local positions within coarse blocks.

\textbf{HQ-only anchor codec.}
This baseline uses the same retained anchors as the proposed method,
but does not use the decoder-side coarse contour. Therefore, it must explicitly encode the ordered anchor coordinates using the first anchor absolute position and subsequent relative coordinate deltas. Since the retained anchors are the same, HQ-only has the same reconstruction quality as the proposed method, but a different bitrate. This comparison directly quantifies the coding gain from decoder-side coarse guidance.

\textbf{JBIG2 on reconstructed contours.}
We also compare with JBIG2, a generic bi-level image codec. For a fair comparison under the same reconstructed geometry, we compress the reconstructed contour maps produced by the proposed method at each threshold using JBIG2.
\subsection{Rate--Distortion Performance}
\label{subsec:rd_results}

\begin{figure}[t]
\centering
\includegraphics[width=0.9\linewidth]{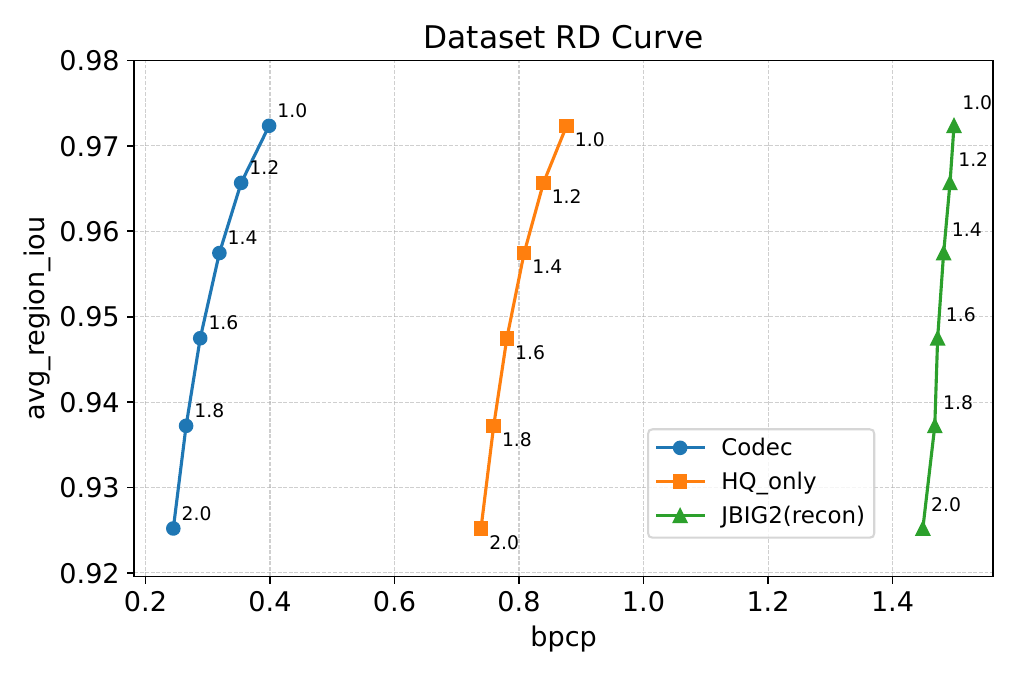}
\caption{
Dataset-level RD performance. The proposed decoder-guided codec achieves a clear left shift compared with HQ-only anchor coding. The JBIG2 points are obtained by compressing the reconstructed contours at each operating point, serving as a generic bi-level coding reference.}
\label{fig:dataset_rd_curve}
\end{figure}

\begin{table}[t]
\centering
\caption{Dataset-level rd results among the proposed codec, HQ-only codec, and JBIG2 on reconstructed contours}
\label{tab:dataset_rd}
\footnotesize
\setlength{\tabcolsep}{6pt}
\renewcommand{\arraystretch}{1}
\begin{tabular}{c c c c c c}
\toprule
\multirow{2}{*}{$\tau$} &
\multirow{2}{*}{IoU$\uparrow$} &
\multicolumn{3}{c}{Bitrate (bpcp)$\downarrow$} &
\multirow{2}{*}{Gain vs. HQ} \\
\cmidrule(lr){3-5}
& & Ours & HQ-only & JBIG2 & \\
\midrule
1.0 & 0.9724 & \textbf{0.3986} & 0.8765 & 1.4987 & 54.5\% \\
1.2 & 0.9657 & \textbf{0.3537} & 0.8392 & 1.4923 & 57.9\% \\
1.4 & 0.9575 & \textbf{0.3188} & 0.8082 & 1.4820 & 60.6\% \\
1.6 & 0.9475 & \textbf{0.2880} & 0.7805 & 1.4724 & 63.1\% \\
1.8 & 0.9372 & \textbf{0.2653} & 0.7593 & 1.4679 & 65.1\% \\
2.0 & 0.9252 & \textbf{0.2448} & 0.7389 & 1.4487 & 66.9\% \\
\bottomrule
\end{tabular}
\end{table}

Fig.~\ref{fig:dataset_rd_curve} shows the RD curves on our dataset. The proposed method consistently achieves lower bpcp than HQ-only at the same region IoU. This confirms that the decoder-available coarse contour provides effective side information: instead of explicitly transmitting the complete ordered anchor coordinate chain, the proposed codec only needs to transmit sparse local refinements along the shared coarse contour order.

Table~\ref{tab:dataset_rd} reports the numerical results. As the
skipping threshold increases from $1.0$ to $2.0$, the proposed method reduces
the bitrate from $0.3986$ bpcp to $0.2448$ bpcp, while the average region IoU decreases from $0.9724$ to $0.9252$. Compared with HQ-only, the proposed codec reduces bitrate by $54.5$\% -- $66.9$\% at identical reconstruction quality. This large saving demonstrates that the coarse contour effectively removes the need
to explicitly encode full anchor ordering and global coordinate deltas. Compared with JBIG2, the proposed method reduces
rate from 1.4987 to 0.3986 bpcp at $\tau=1.0$, i.e., a 3.76$\times$ saving. The advantage becomes larger at lower-rate operating points.

%The results show our anchor skipping provides a smooth and controllable rate- distortion trade-off, and the gap between the proposed method and HQ-only reveals the practical value of decoder-side guidance. 

%Increasing $\tau$ produces a monotonic bitrate reduction, while the region IoU degrades gradually rather than abruptly. This indicates that the skipped anchors are mostly redundant for preserving coarse object geometry.

%Second, . HQ-only must transmit the ordered anchor coordinate chain explicitly, whereas the proposed codec inherits ordering and connectivity from the coarse contour. Therefore, the proposed method is especially efficient when the decoder already has access to a low-quality reference from which coarse object boundaries can be extracted. This setting is common in super-resolution, base-layer reconstruction, and other decoder-guided visual reconstruction pipelines.

\subsection{Ablation}
\label{subsec:ablation}

We further conduct ablation studies to analyze the effect of the coarse block size and the contribution of entropy coding. 

\textbf{Effect of block size.} Table~\ref{tab:blocksize_ablation} compares block sizes $s=3,4,6$ on our dataset. A smaller block $s=3$ lowers the bitrate by about $6.1\%$ compared with $s=4$, but is more sensitive to aggressive skipping, leading to faster IoU
degradation at large $\tau$. A larger block $s=6$ improves the average region IoU by about $0.008$ over the default setting, but increases the codec bitrate by about $9.5\%$. Larger blocks provide a more stable coarse structure and preserve the reconstructed contour better under anchor skipping, but require a larger local-position alphabet and result in higher coding cost.  

\begin{table}[t]
\centering
\caption{Ablation on coarse block size. Each entry reports bpcp / IoU, illustrating the rate--accuracy trade-off of different coarse-grid resolutions}
\label{tab:blocksize_ablation}
\footnotesize
\setlength{\tabcolsep}{6pt}
\renewcommand{\arraystretch}{1}
\begin{tabular}{c c c c}
\toprule
\multirow{2}{*}{$\tau$} &
\multicolumn{3}{c}{Block size $s$: bpcp$\downarrow$ / IoU$\uparrow$} \\
\cmidrule(lr){2-4}
& $s=3$ & $s=4$ & $s=6$ \\
\midrule
1.0 & 0.3753 / 0.9704 & 0.3986 / 0.9724 & 0.4212 / 0.9714 \\
1.2 & 0.3300 / 0.9601 & 0.3537 / 0.9657 & 0.3829 / 0.9679 \\
1.4 & 0.2967 / 0.9476 & 0.3188 / 0.9575 & 0.3490 / 0.9630 \\
1.6 & 0.2694 / 0.9331 & 0.2880 / 0.9475 & 0.3187 / 0.9573 \\
1.8 & 0.2503 / 0.9177 & 0.2653 / 0.9372 & 0.2951 / 0.9513 \\
2.0 & 0.2319 / 0.8999 & 0.2448 / 0.9252 & 0.2726 / 0.9442 \\
\bottomrule
\end{tabular}
\end{table}

\begin{table}[t]
\centering
\caption{Ablation on entropy coding. All variants have the same IoU at each $\tau$; only bpcp changes.}
\label{tab:entropy_ablation}
\footnotesize
\setlength{\tabcolsep}{5pt}
\renewcommand{\arraystretch}{1}
\begin{tabular}{c c c c c c c}
\toprule
\multirow{2}{*}{$\tau$} &
\multirow{2}{*}{IoU$\uparrow$} &
\multicolumn{4}{c}{Bitrate (bpcp)$\downarrow$} &
\multirow{2}{*}{Gain} \\
\cmidrule(lr){3-6}
& & No EC & Rice only & Huff. only & Full & \\
\midrule
1.0 & 0.9724 & 0.6588 & 0.4059 & 0.6458 & \textbf{0.3986} & 39.5\% \\
1.2 & 0.9657 & 0.5657 & 0.3602 & 0.5592 & \textbf{0.3537} & 37.5\% \\
1.4 & 0.9575 & 0.4919 & 0.3240 & 0.4867 & \textbf{0.3188} & 35.2\% \\
1.6 & 0.9475 & 0.4272 & 0.2921 & 0.4231 & \textbf{0.2880} & 32.6\% \\
1.8 & 0.9372 & 0.3790 & 0.2684 & 0.3759 & \textbf{0.2653} & 30.0\% \\
2.0 & 0.9252 & 0.3361 & 0.2473 & 0.3335 & \textbf{0.2448} & 27.2\% \\
\bottomrule
\end{tabular}
\end{table}

\textbf{Effect of entropy coding.} Table~\ref{tab:entropy_ablation} isolates the bitrate contribution of entropy coding; all variants share the same reconstruction and therefore the same IoU. Rice coding provides the dominant gain because retained-index differences are concentrated near small positive values, In contrast, anchor local positions are more evenly distributed within a coarse block, so the global Huffman codebook brings a smaller gain. The full entropy coding scheme
reduces bpcp by up to 39.5\% and by 33.7\% on average over fixed-length syntax.

\section{Conclusion}
\label{sec:conclusion}

In this paper, we propose a decoder-guided coarse-to-fine contour coding framework for efficient lossy contour
compression. By exploiting a decoder-available coarse contour as geometric side information, the proposed method avoids explicitly transmitting full contour, and instead codes only sparse anchor refinements with controllable rate–distortion trade-offs. Experiments on SportsSloMo demonstrate that the proposed method substantially reduces bitrate compared with both HQ-only anchor coding and JBIG2 while maintaining high contour accuracy. These results suggest that decoder-guided geometric refinement is a promising direction for compact contour representation in bandwidth-constrained visual reconstruction pipelines. Future work will integrate the proposed compressed-contour representation into
end-to-end SR/VSR and generation frameworks to evaluate its impact on downstream perceptual and task-specific reconstruction quality.

% \begin{acks}
% %% Acknowledgments go here. For example:
% %% This work was supported by ...
% \end{acks}
\FloatBarrier
\bibliographystyle{IEEEtran}
\bibliography{references}

\end{document}